\documentclass[iop]{emulateapj}

\usepackage{multirow}

\newcommand{\nustar} {{\it NuSTAR}}

\newcommand{\cmsq} {cm$^{-2}$}
\newcommand{\nh} {$N_{\rm{H}}$}
\newcommand{\lx} {$L_{\rm{X}}$}

\newcommand{\chisq} {$\chi^2$}

\newcommand{\hb}{{\rm{H$\beta$}}}

\newcommand{\ergs}{\mbox{\thinspace erg\thinspace s$^{-1}$}}

\newcommand{\lbol} {$L_{\rm Bol}$}
\newcommand{\kbol} {$\kappa_{\rm Bol}$}
\newcommand{\mbh} {$M_{\rm BH}$}

\newcommand{\lamedd} {$\lambda_{\rm Edd}$}

\newcommand{\msol} {$M_{\odot}$}
\newcommand{\lsol} {$L_{\odot}$}

\shorttitle{A growth-rate indicator for CTAGN}
\shortauthors{Brightman et al.}

\begin{document}

\title{A growth-rate indicator for Compton-thick active galactic nuclei}

\author{M. Brightman$^{1}$, A. Masini$^{2,3}$, D. R. Ballantyne$^{4}$, M. Balokovi\'{c}$^{1}$, W. N. Brandt$^{5}$, C.-T. Chen$^{5}$, A. Comastri$^{2}$, D. Farrah$^{6}$, P. Gandhi$^{7}$, F. A. Harrison$^{1}$, C. Ricci$^{8}$, D. Stern$^{9}$, D. J. Walton$^{9,1}$}

\affil{$^{1}$Cahill Center for Astrophysics, California Institute of Technology, 1216 East California Boulevard, Pasadena, CA 91125, USA\\
$^{2}$INAF Osservatorio Astronomico di Bologna, via Ranzani 1, I-40127 Bologna, Italy\\
$^{3}$Dipartimento di Fisica e Astronomia (DIFA), Universit\'{a} di Bologna, viale Berti Pichat 6/2, 40127 Bologna, Italy\\
$^{4}$Center for Relativistic Astrophysics, School of Physics, Georgia Institute of Technology, Atlanta, GA 30332, USA\\
$^{5}$Department of Astronomy and Astrophysics, The Pennsylvania State University, University Park, PA 16802, USA\\
$^{6}$Department of Physics, Virginia Tech, Blacksburg, VA 24061, USA\\
$^{7}$Department of Physics and Astronomy, University of Southampton, Highfield, Southampton SO17 1BJ, UK\\
$^{8}$Instituto de Astrof\'{i}sica, Facultad de F\'{i}sica, Pontificia Universidad Cat\'{o}lica de Chile, Casilla 306, Santiago 22, Chile\\
$^{9}$Jet Propulsion Laboratory, California Institute of Technology, Pasadena, CA 91109, USA\\}

\begin{abstract}
Due to their heavily obscured central engines, the growth rate of Compton-thick (CT) active galactic nuclei (AGN) is difficult to measure. A statistically significant correlation between the Eddington ratio, \lamedd, and the X-ray power-law index, $\Gamma$, observed in unobscured AGN offers an estimate of their growth rate from X-ray spectroscopy (albeit with large scatter). However, since X-rays undergo reprocessing by Compton scattering and photoelectric absorption when the line-of-sight to the central engine is heavily obscured, the recovery of the intrinsic $\Gamma$ is challenging. Here we study a sample of local, predominantly Compton-thick megamaser AGN, where the black hole mass, and thus Eddington luminosity, are well known. We compile results on X-ray spectral fitting of these sources with sensitive high-energy ($E>10$ keV) \nustar\ data, where X-ray torus models which take into account the reprocessing effects have been used to recover the intrinsic $\Gamma$ values and X-ray luminosities, \lx. With a simple bolometric correction to \lx\ to calculate \lamedd, we find a statistically significant correlation between $\Gamma$ and \lamedd\ ($p=0.007$). A linear fit to the data yields $\Gamma=( 0.41\pm 0.18)$log$_{10}$\lamedd$+( 2.38\pm 0.20)$, which is statistically consistent with results for unobscured AGN. This result implies that torus modeling successfully recovers the intrinsic AGN parameters. Since the megamasers have low-mass black holes (\mbh$\approx10^{6}-10^{7}$ \msol) and are highly inclined, our results extend the $\Gamma$-\lamedd\ relationship to lower masses and argue against strong orientation effects in the corona, in support of AGN unification. Finally this result supports the use of $\Gamma$ as a growth-rate indicator for accreting black holes, even for Compton-thick AGN.

\end{abstract}

\keywords{galaxies -- black hole physics -- masers -- galaxies: nuclei -- galaxies: Seyfert}

\section{Introduction}

Determining the growth rates of active galactic nuclei (AGN) is important for understanding the build up of supermassive black holes. The key parameter to describe black hole growth is the Eddington ratio, \lamedd. This is defined by the ratio of the bolometric luminosity of the AGN, \lbol, to the Eddington luminosity, $L_{\rm Edd}$ (i.e. \lamedd$\equiv$\lbol/$L_{\rm Edd}$). \lbol\ is related to the mass accretion rate onto the black hole, $\dot{m}$, via the accretion efficiency, $\eta$, by $L_{\rm Bol}=\eta\dot{m}c^{2}$. $L_{\rm Edd}$ is the theoretical maximal luminosity (although observed to be exceeded in some sources e.g. \cite{lanzuisi16}) achieved via accretion when accounting for radiation pressure, and is dependent on the black hole mass ($L_{\rm Edd}=4\pi$G\mbh\ $m_{\rm p}c/\sigma_{\rm T}\simeq1.26\times10^{38}$(\mbh/\msol) \ergs).

For unobscured AGN, \lamedd\ is determined from the intrinsic disk emission observed in the optical/UV, from which \lbol\ can be calculated, and \mbh\ that is estimated from measurements of the broad emission lines which trace the motions of gas close to the black hole \citep[e.g.][]{shen13,peterson14}. 

In obscured AGN however, the intrinsic disk emission is completely extinguished by intervening material, and the broad line region is obscured from view, so \lamedd\ is difficult to measure in these systems and must be estimated from indirect methods. For example, the observed relationship between the stellar velocity dispersion in the bulge of the galaxy and the black hole mass is often used to estimate \mbh. However, this relationship has a large intrinsic scatter in it, especially at low masses \citep[e.g.][]{greene10,laesker16}. It is therefore important to have as many indirect methods as possible for estimating \lamedd\ for both unobscured and obscured AGN.

Studies of the X-ray emission of AGN have found that \lamedd\ is strongly correlated with the X-ray spectral index, $\Gamma$, in the range $0.01\lesssim$\lamedd$\lesssim1$ \citep[e.g.,][]{shemmer06,shemmer08,risaliti09,jin12,brightman13}. $\Gamma$ depends on both the electron temperature and optical depth to Compton scattering in the hot corona \citep{rybicki86,haardt93,fabian15} that up-scatters the optical/UV emission from the accretion disk \citep[e.g.][]{shakura73}. This relationship is thought to arise due to higher \lamedd\ systems cooling their coronae more effectively than lower \lamedd\ through enhanced optical/UV emission. 

The observed relationship between $\Gamma$ and \lamedd\ suggests that a measurement of $\Gamma$ could be used to estimate \lamedd. This would be particularly useful for heavily obscured AGN due to the fact that \lamedd\ is, as mentioned, difficult to measure for such systems. However, this has its own challenges, since X-rays are also absorbed in these sources and at large column densities (\nh$\sim10^{24}$ \cmsq) X-rays undergo Compton-scattering within the obscuring medium, which modifies their trajectory and energy. Nonetheless, up to \nh$\sim10^{25}$ \cmsq\ and at high energy ($E>10$ keV) absorption is negligible, and furthermore spectral models exist that take these effects into account, assuming a torus geometry of the obscuring medium, e.g. {\tt mytorus} \citep{murphy09} and {\tt torus} \citep{brightman11}. In order to recover the intrinsic $\Gamma$ using these models, broadband X-ray spectral measurements, especially above 10 keV where the scattering dominates, are required. \nustar\ \citep{harrison13}, with its sensitivity at these energies, is the ideal instrument with which to uncover the intrinsic X-ray emission from heavily obscured AGN and since its launch in 2012 has amassed a large archive of data on these sources \citep[e.g.][]{puccetti14,arevalo14,balokovic14,gandhi14,bauer15,brightman15,koss15,annuar15,rivers15b,marinucci16,ricci16}

In this work our goal is to examine the relationship between $\Gamma$ and \lamedd\ for heavily obscured AGN to test if it is consistent with the results from unobscured AGN. This will reveal how well X-ray spectral modeling with the X-ray torus models recovers the intrinsic AGN parameters, or if orientation effects in the corona are present, related to AGN unification, and show if $\Gamma$ can be used as a \lamedd\ indicator for these heavily obscured systems. 

This requires a sample of heavily obscured AGN where the black hole mass has been measured reliably and broadband X-ray spectra are available. The most robust black hole mass measurements for obscured AGN come from disk water megamasers \citep[see][for a review]{lo05}, where the Keplerian motion of the masing material reveals the mass within \citep[e.g.,][]{greenhill96}. Due to the edge-on geometry of the medium required to produce masing emission, a high fraction of megamasers are heavily obscured AGN \citep{zhang06,masini16}, making megamasers particularly well suited to our study.

Furthermore, megamasers are of interest since they are at the low-mass end of the supermassive black hole mass distribution, having a mass range of \mbh$\approx10^6-10^7$ \msol. Previous analyses of the $\Gamma$-\lamedd\ relationship have concentrated on samples where the black hole mass has been measured from optical broad line fitting \citep[e.g.,][]{brightman13} with \mbh$\approx10^7-10^9$ \msol. More recently, lower-mass black holes (\mbh$\sim10^6$ \msol) have been investigated \citep[e.g. selected via their rapid X-ray variability,][]{kamizasa12}, where it has been found that they are not fully consistent with the results from higher mass \citep{ai11,ho16}. The megamaser AGN thus give us the opportunity to further assess the validity of the relationship in this low-mass regime with a different sample selection.

We describe our sample and its selection in Section \ref{sec_sample}, give our results in Section \ref{sec_results} and discuss and conclude in Section \ref{sec_disc}.

\section{Megamaser sample}
\label{sec_sample}

There are $\sim20$ sources where megamaser emission has been used to measure black hole mass \citep[e.g.][]{kuo11}. For our analysis, we require results from sensitive broadband X-ray spectral data, especially above 10~keV where Compton scattering effects dominate. For this reason we compile \nustar\ results on the megamaser AGN. This was done recently by \cite{masini16} who compiled and analyzed X-ray spectral information of megamaser AGN in order to study the connection between the masing disk and the torus. These AGN include well-studied sources that have been the subject of detailed spectral analysis of \nustar\ data, such as Circinus \citep{arevalo14}, NGC~4945 \citep{puccetti14}, NGC~1068 \citep{bauer15,marinucci16}, and NGC~3393 \citep{koss15}, as well as samples of sources such as IC~2560, NGC~1368, and NGC~3079 \citep{balokovic14,brightman15}. 

In all of these studies, the {\tt mytorus} and {\tt torus} models were used to obtain the intrinsic $\Gamma$ and 2$-$10~keV luminosities, \lx, correcting for columns of 10$^{23}-10^{26}$~\cmsq. In most studies of the megamaser AGN listed above, both {\tt mytorus} and {\tt torus} models were fitted, with generally good agreement between spectral parameters \citep[for a direct comparison see][]{brightman15}. For our study, we take the results on $\Gamma$ and \lx\ from the model that the original authors found to be the best fitting one.  

In order to test the $\Gamma$-\lamedd\ relationship for the megamaser AGN, we require good constraints on $\Gamma$, thus we exclude sources where the uncertainty on $\Gamma$ is $>0.25$, which excludes NGC 1386 and NGC 2960 from our sample. Our final sample consists of nine AGN. For NGC~4945, \cite{puccetti14} present a flux resolved analysis of the source, whereby they investigated the variation of $\Gamma$ with the source luminosity (and hence \lamedd), which is of particular interest here, so we include those individual results here, giving us 12 separate measurements of $\Gamma$ for the sample. With the exception of NGC~4388 \citep[\nh$=4\times10^{23}$ \cmsq,][]{masini16}, this sample consists wholly of Compton-thick (\nh$\geq1.5\times10^{24}$ \cmsq) AGN.

With black hole masses from the megamasers and good measurements of $\Gamma$, the final ingredient required for our investigation is \lbol, needed to calculate \lamedd. Since the X-ray spectral modeling also yields intrinsic $2-10$ keV luminosities, \lx, for our sample, the simplest approach is to apply a bolometric correction, \kbol, to \lx. Several works have presented results on \kbol, reporting that it is an increasing function of \lbol\ \citep[e.g.][]{marconi04,hopkins07}, or that it is a function of \lamedd\ \citep{vasudevan07}. From a large X-ray selected sample in $XMM$-COSMOS, \cite{lusso12} confirm that \kbol\ is a function of both \lbol\ and \lamedd. Given the relatively low X-ray luminosities of our sample (\lx$\sim10^{42}-10^{43}$ \ergs) which correspond to bolometric luminosities of $\sim10^{10}-10^{11}$ \lsol, the results from \cite{lusso12} show that \kbol$=10$ would be appropriate for these sources. Thus for our initial investigation we calculate \lamedd\ in this way. 

The uncertainty on \lamedd\ is propagated from the uncertainty in \mbh\ and in \lx\ by adding them in quadrature. For \mbh\ the uncertainty is typically $\sim5$\% or higher. For \lx\ we assume a systematic 25\% uncertainty to account for uncertainties in the flux from spectral modeling and any uncertainty in the distance to the source, which for these nearby galaxies can be non-negligible. We explore the effect of calculating \lbol\ from a bolometric correction on our results later in the paper, as well as the use of \lbol\ estimated from multiwavelength data. The properties of our sample are summarized in Table \ref{tab_sample}.

\begin{table*}
\centering
\caption{Properties of the \nustar\ megamaser sample}
\label{tab_sample}
\begin{center}
\begin{tabular}{l c c c c c c c l}
\hline
AGN Name & Redshift & \mbh$/10^6$ \msol & log$_{10}$(\lx/\ergs) & $\Gamma$ & \nh/$10^{24}$ \cmsq 	& \lamedd & References \\
(1) & (2) & (3) & (4) & (5) & (6) & (7) & (8) \\
\hline
NGC 1068&0.0038&  8.0$\pm$0.3&43.34& 2.10$\pm$0.07&$5.0^{+4.2}_{-1.9}$& 0.210$\pm$0.053&c, l\\
NGC 1194&0.0136& 65.0$\pm$3.0&42.78& 1.59$\pm$0.15&$1.4^{+0.3}_{-0.2}$& 0.007$\pm$0.002&f, m\\
NGC 2273&0.0061&  7.5$\pm$0.4&43.11& 2.10$\pm$0.10&$>7.3$& 0.132$\pm$0.034&f, m\\
NGC 3079&0.0037&  2.4$^{+2.4}_{-1.2}$&41.53& 1.86$\pm$0.25&$1.84\pm0.32$& 0.011$\pm$0.009&d, j\\
NGC 3393&0.0125& 31.0$\pm$2.0&43.40& 1.82$\pm$0.09&$2.2\pm0.4$& 0.062$\pm$0.016&e, k\\
NGC 4388&0.0084&  8.5$\pm$0.2&42.59& 1.65$\pm$0.08&$0.44\pm0.06$& 0.035$\pm$0.009&f, m\\
NGC 4945 (L)&\multirow{4}{*}{0.0019}&\multirow{4}{*}{1.4$\pm$0.7}&42.09& 1.77$\pm$0.09&$3.5\pm0.2$& 0.068$\pm$0.038&\multirow{4}{*}{a, i}\\
NGC 4945 (M)&&&42.39& 1.88$\pm$0.05&$3.6\pm0.1$& 0.135$\pm$0.075&\\
NGC 4945 (H)&&&42.62& 1.95$\pm$0.04&$3.6\pm0.1$& 0.229$\pm$0.128&\\
NGC 4945 (SH)&&&42.74& 1.96$\pm$0.07&$3.5\pm0.1$& 0.302$\pm$0.169&\\
IC 2560&0.0098&  3.5$\pm$0.5&42.90& 2.50$\pm$0.20&$>13$& 0.175$\pm$0.050&g, j\\
Circinus&0.0014&  1.7$\pm$0.3&42.50& 2.27$\pm$0.05&$8.9\pm1.2$& 0.143$\pm$0.044&b, h\\

\hline
\end{tabular}
\tablecomments{Column (1) lists the name of the megamaser AGN, where four different entries for NGC~4945 are given when it was observed at low (L), medium (M), high (H) and super-high (SH) flux levels \citep[see][]{puccetti14}. Column (2) gives the redshift of the source, column (3) lists the black hole mass in units of 10$^6$ \msol, column (4) gives the logarithm of the intrinsic 2-10 keV luminosity of the AGN determined through spectral modeling, column (5) gives $\Gamma$, column (6) gives the \nh\ in 10$^{24}$ \cmsq, column (7) shows the Eddington ratio, \lamedd\ given a bolometric correction of 10 to \lx. In column (8) we give the reference for the black hole mass - a. \cite{greenhill97}, b. \cite{greenhill03}, c. \cite{lodato03}, d. \cite{kondratko05}, e. \cite{kondratko08}, f. \cite{kuo11}, g. \cite{yamauchi12}, and the X-ray spectral information - h. \cite{arevalo14}, i. \cite{puccetti14}, j. \cite{brightman15}, k. \cite{koss15}, l. \cite{bauer15}, m. \cite{masini16}.}

\end{center}
\end{table*}

\section{Are the heavily obscured megamasers consistent with unobscured AGN?}
\label{sec_results}

\begin{figure}
\begin{center}
\includegraphics[width=90mm]{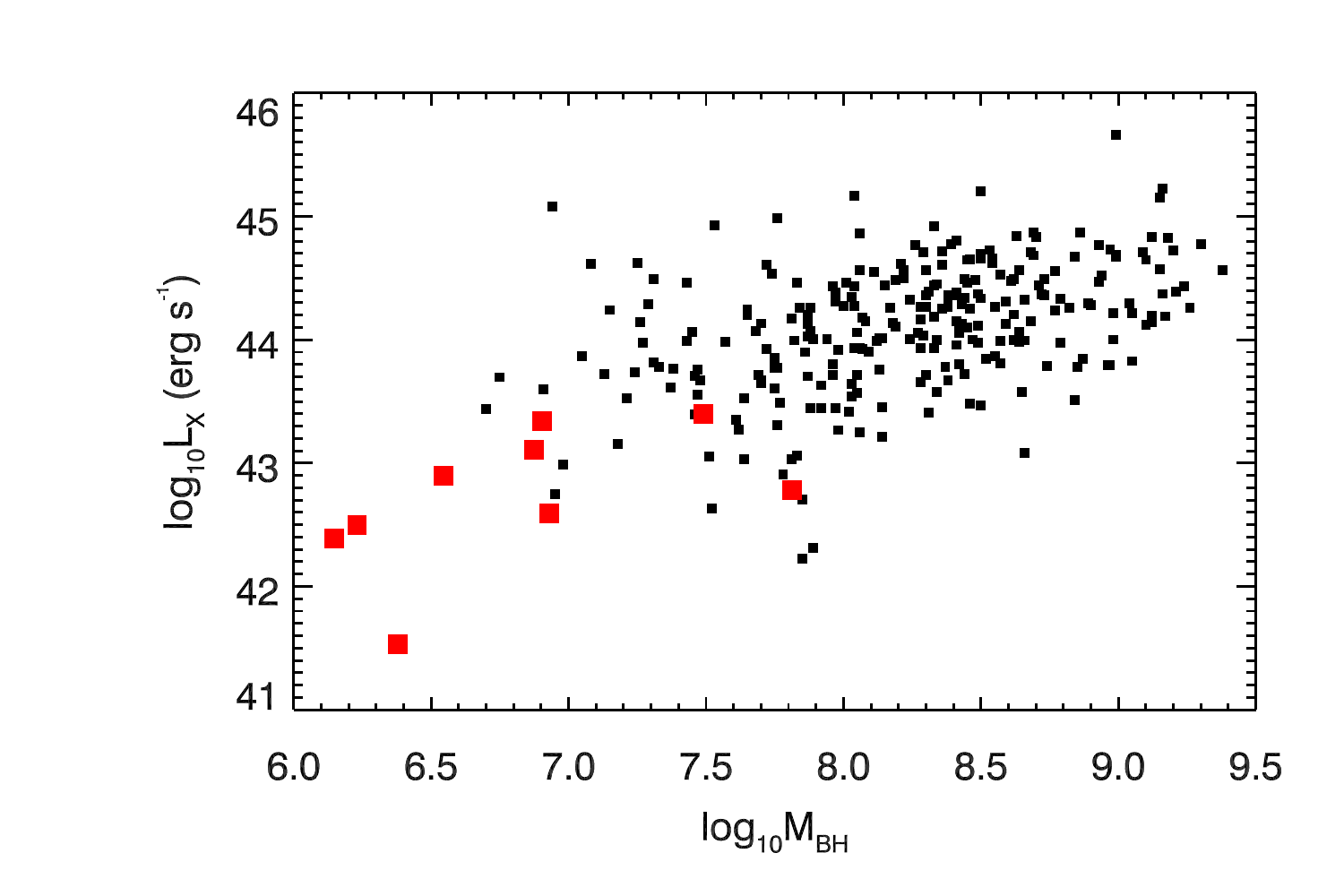}
\caption{The black hole mass and 2$-$10 keV luminosity distributions of the megamaser AGN (red points) compared to the broad-lined AGN sample of B13 (black points).}
\label{fig_mbh_lx}
\end{center}
\end{figure}

For our comparison of the $\Gamma$-\lamedd\ relationship for megamaser AGN with unobscured AGN we use the sample of \cite{brightman13} (B13), who studied a sample of 69 unobscured AGN in the Cosmic Evolution Survey \citep[COSMOS,][]{scoville07} and Extended Chandra Deep Field-South \citep[E-CDF-S,][]{lehmer05} survey up to $z\sim2$ with black hole masses measured from optical broad line measurements. B13 fit the X-ray spectra of their sources in the $2-10$ keV range with a simple power-law model. We plot the \mbh\ and \lx\ distributions of the megamaser sample in Figure \ref{fig_mbh_lx} compared to the sample of B13. This shows that the megamaser sample extends the study of the $\Gamma$-\lamedd\ relationship to lower black hole masses.

We first test for the significance of a correlation between $\Gamma$ and \lamedd\ in the megamaser AGN with a Spearman rank correlation test. This yields $r_{\rm S}=0.73$ and $p=0.007$, where $r_{\rm S}$ is the Spearman rank correlation coefficient and $p$ is the probability of obtaining the absolute value of $r_{\rm S}$ at least as high as observed, under the assumption of the null hypothesis of zero correlation. The small value of $p$ indicates a significant correlation as observed in samples of unobscured AGN.

\begin{figure}
\begin{center}
\includegraphics[width=90mm]{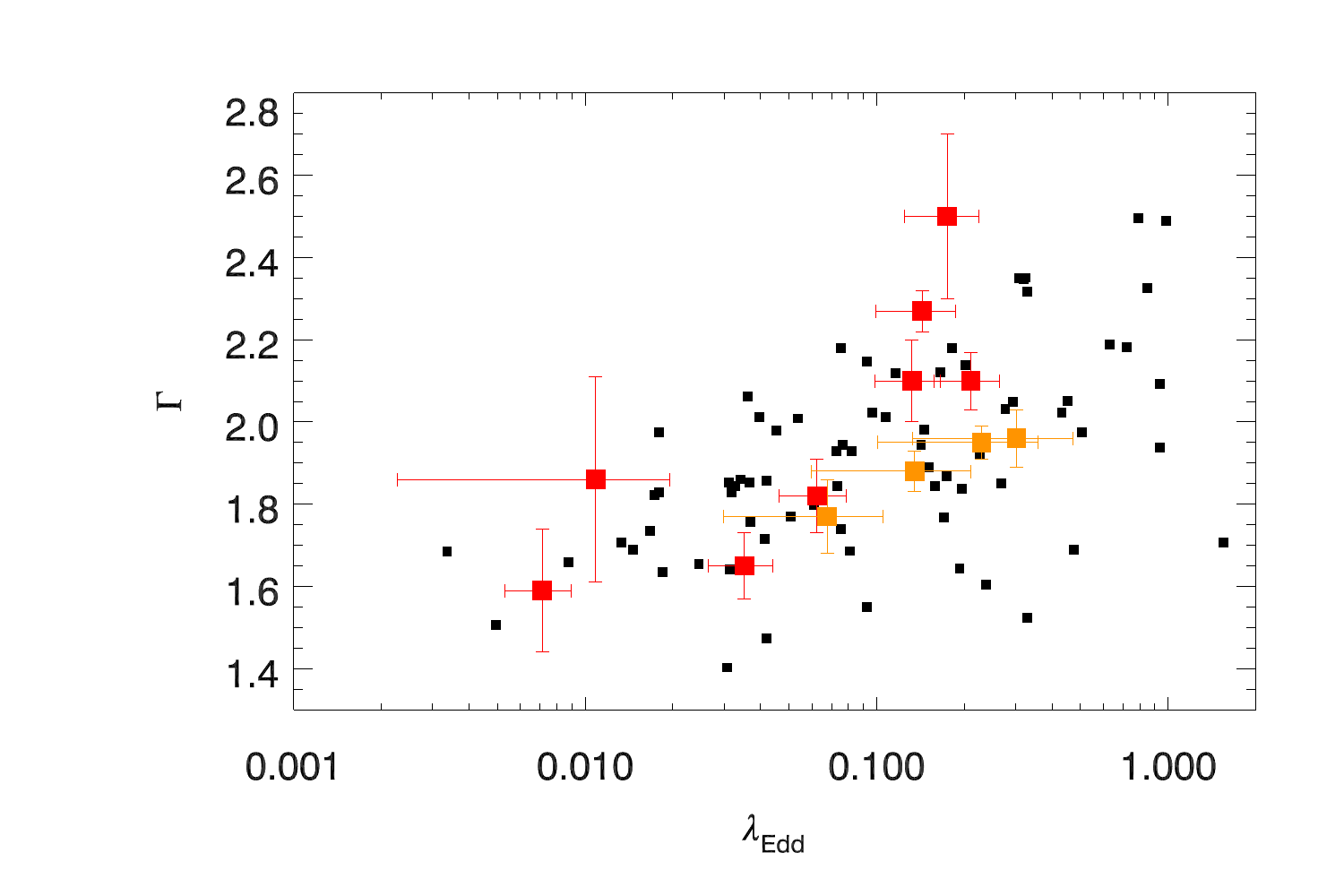}
\caption{The $\Gamma$ and \lamedd\ distributions of the megamaser AGN (red points, with the multiple measurements of NGC~4945 in orange) compared to the broad-lined AGN sample of B13 (black points).}
\label{fig_lam_gam}
\end{center}
\end{figure}

We present a comparison of the distribution of $\Gamma$ and \lamedd\ for the megamaser AGN to the unobscured AGN in Figure \ref{fig_lam_gam}. This shows that the two AGN samples occupy the same locus, given the measurement uncertainties, suggesting that they are drawn from the same underlying population. We test this quantitatively by fitting a linear regression to the megamaser AGN data, as done for the unobscured AGN, and compare the results. So that a direct comparison can be made, we use the {\sc idl} function {\sc linfit} as done by B13, which fits the paired data \{\lamedd$_{i}$, $\Gamma_{i}$\} to the linear model, $\Gamma = m$log$_{10}$\lamedd$+c$, by minimizing the \chisq\ error statistic. The measurement uncertainties on $\Gamma$ are used to compute the \chisq\ statistic (the uncertainty on \lamedd\ is neglected). 

The result from the linear fitting yields $\Gamma=( 0.31\pm 0.07)$log$_{10}$\lamedd$+( 2.24\pm 0.06)$, where \chisq=59.9 for 10 degrees of freedom.  The same fit to the sample of B13 gave $\Gamma=(0.32\pm0.05)$log$_{10}$\lamedd$+(2.27\pm0.06)$. Both the slopes and offsets of the linear relationships are in very good agreement. 

Other results on unobscured AGN from \cite{shemmer08} (S08) and \cite{risaliti09} (R09) found similar values for the slope of the relationship, $0.31\pm0.01$ and $0.31\pm0.06$, respectively; thus, the results from the megamaser AGN are also consistent with these results. As for the offsets, S08 measure $2.11\pm0.01$ and R09 measure $1.97\pm0.02$. However, R09 calculate their linear fit with log$_{10}$\lamedd$=-1$ as their reference point, rather than 0 as we have done here, which corresponds to $c=2.28$ with log$_{10}$\lamedd$=0$ as the reference point. Thus the offsets are consistent within $\sim1-2\sigma$. 

Other authors have, however, found steeper slopes in the relationship. \cite{jin12} find a slope of 0.58 from a sample of unobscured nearby type 1 AGN, while \cite{keek16} find that the slope is 0.54 when fitting for $\Gamma$ vs. \lamedd\ in varying states of Mrk~335. These slopes are similar to that found by R09 for black hole masses based on the \hb\ line only (0.58). Some of this disagreement appears to be due to the different statistical analyses used. \cite{jin12} suggest that \chisq\ minimization may not be appropriate for quantifying this relation because it can be biased by small measurement errors in $\Gamma$ for individual sources. The \chisq\ normalization also does not take into account uncertainties in \lamedd\ or any intrinsic scatter. Indeed the \chisq/DOF of 59.9/10 that we find from this indicates significant scatter is indeed present. 

\cite{kelly07} presented a Bayesian method to account for measurement errors in linear regression of astronomical data, {\sc linmix\_err}, which also takes into account uncertainties in the independent variable and allows for intrinsic dispersion in the regression. Applying this code to our data yields $\Gamma=( 0.41\pm 0.18)$log$_{10}$\lamedd$+( 2.38\pm 0.20)$ with an intrinsic scatter of 0.19$\pm$0.19. While the slope is steeper compared to the \chisq\ minimization result, the uncertainties are larger due to the inclusion of the \lamedd\ uncertainties. The slopes of the \chisq\ minimization and Bayesian methods are within 1-$\sigma$ of each other as well as with the slopes from the unobscured AGN. This is likewise true of the offset, which is slightly higher with respect to the \chisq\ minimization result, but the larger uncertainty makes it consistent within 2-$\sigma$ of all the results on the unobscured AGN.

\begin{figure}
\begin{center}
\includegraphics[width=90mm]{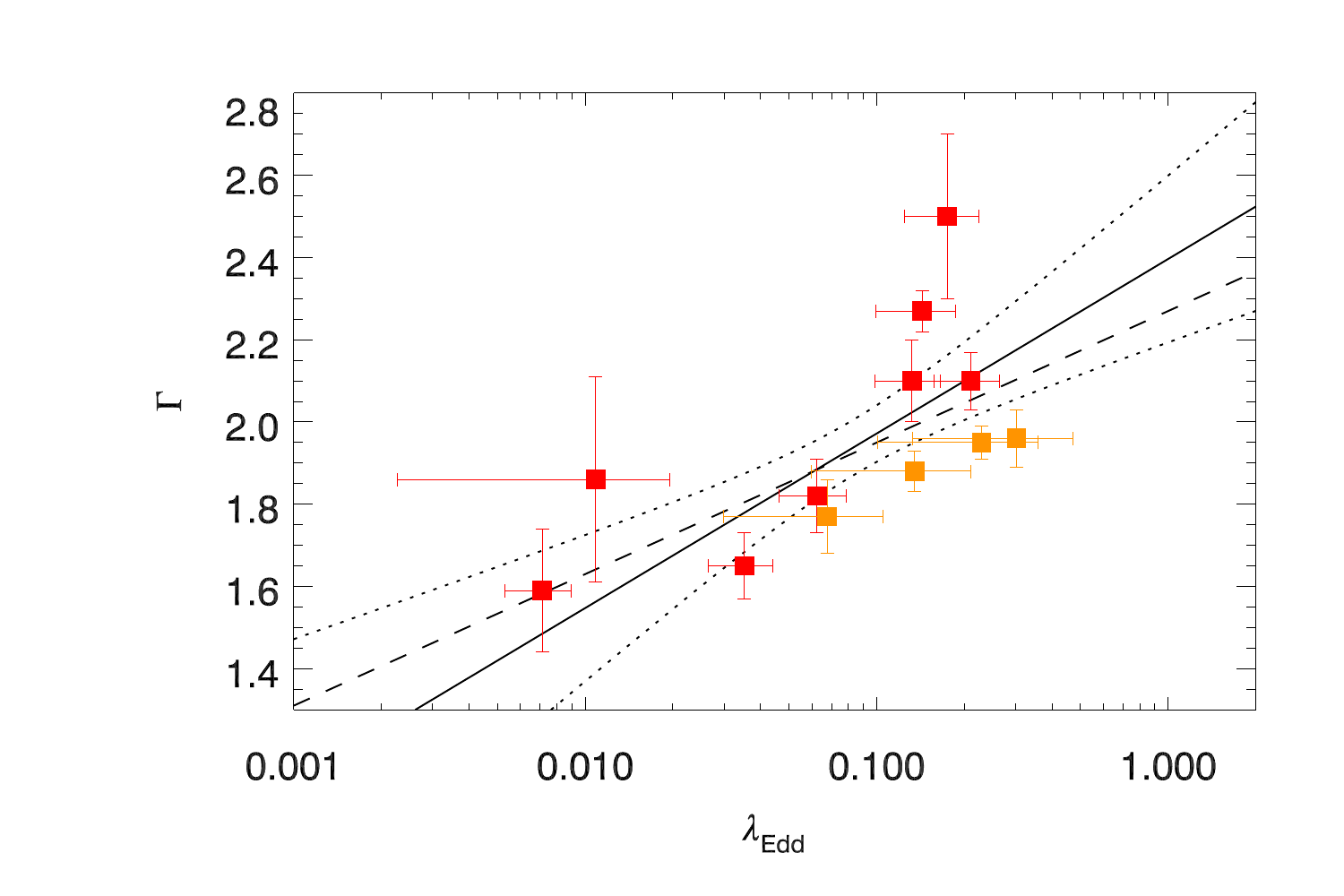}
\caption{A linear-regression fit to the $\Gamma$ and \lamedd\ distributions of the megamaser AGN yields $\Gamma=( 0.41\pm 0.18)$log$_{10}$\lamedd$+( 2.38\pm 0.20)$, shown by the solid black line. The dotted lines mark the upper and lower 1-$\sigma$ confidence limits given the uncertainties on the slope and offset of the linear relationship. The dashed line shows the linear relationship derived from unobscured AGN from B13 demonstrating very good agreement between the two, given the uncertainties. As for Figure \ref{fig_lam_gam}, the data points are plotted in red, with the multiple measurements of NGC~4945 highlighted in orange}
\label{fig_lam_gam_lin}
\end{center}
\end{figure}

We plot the data with the result of the linear fit with the Bayesian method along with the upper and lower 1-$\sigma$ confidence bounds in Figure \ref{fig_lam_gam_lin}. The confidence bounds have been determined from a draw from the posterior distribution of the slope and offset parameters. 

\subsection{How does the calculation of \lbol\ affect our results?}

The largest source of systematic uncertainty in these results come from our estimation of \lbol\ and consequently \lamedd, which we calculate given a bolometric correction, \kbol\ to the intrinsic $2-10$ keV luminosity, \lx. Our initial choice of \kbol\ comes from the relatively low \lx\ of our sample, for which the results from \cite{lusso12} show that \kbol$\approx10$. Firstly, \cite{lusso12} find that \kbol\ is in increasing function of \lbol. In the range of \lx\ we consider here, the function is relatively flat, which justifies our use of a constant value. However, we check our results using the functional form of \kbol\ against luminosity presented by \cite{lusso12} for type 2 AGN from their combined spectroscopic and photometric redshift sample. We find no change in the resulting slope and offset in the $\Gamma$-\lamedd\ relationship from this.

In addition to this, the relationship between \kbol\ and \lbol\ has a large intrinsic scatter, with \kbol\ greater than 100 inferred for the most luminous sources. We therefore examine the effect of different choices of \kbol\ on our results, testing \kbol=$5, 10, 20, 30$ and 50.  Table \ref{tab_kbol} presents the results from this analysis, which shows how the the linear fit to the data $\Gamma = m$log$_{10}$\lamedd$+c$ with \chisq\ minimization is affected. As expected, the choice of a constant \kbol\ does not effect the slope of this relationship since increasing \kbol\ systematically increases \lamedd. The effect of increasing \kbol\ is to decrease the offset of the relationship from $2.33\pm 0.08$ for \kbol$=5$ to $ 2.02\pm 0.02$ for \kbol$=50$. 

We also test the case that \kbol\ is dependent on \lamedd. \cite{vasudevan07} find a transitional region at \lamedd$\sim$0.1, below which \kbol$=15-25$, and above which it is $40-70$. To apply this, we apply an initial \kbol\ of 20 to the sample. For sources where \lamedd$>0.1$ results from this, we recalculate \lbol\ using \kbol$=40$. The result of this is to flatten out the linear relationship, such that the slope becomes $ 0.26\pm 0.05$ with the offset more consistent with higher \kbol\ values ($ 2.05\pm 0.03$).

\begin{table}
\centering
\caption{Investigating the choice of bolometric correction.}
\label{tab_kbol}
\begin{center}
\begin{tabular}{l c c}
\hline
Result & $m$ & $c$ \\
(1) & (2) & (3)\\
\hline
\kbol$=5$			& $ 0.31\pm 0.07$	& $ 2.33\pm 0.08$	 \\
\kbol$=10$		& $ 0.31\pm 0.07$	& $ 2.24\pm 0.06$	 \\
\kbol$=20$		& $ 0.31\pm 0.07$	& $ 2.15\pm 0.04$	\\
\kbol$=30$		& $ 0.31\pm 0.07$	& $ 2.09\pm 0.03$	\\
\kbol$=50$		& $ 0.31\pm 0.07$	& $ 2.02\pm 0.02$	 \\
\kbol$=20$ (\lamedd$\leq0.1$)& \multirow{2}{*}{$ 0.26\pm 0.05$} & \multirow{2}{*}{$ 2.05\pm 0.03$} \\
\kbol$=40$ (\lamedd$>0.1$) \\
\hline
S08			& $0.31\pm0.01$	& $2.11\pm0.01$ \\
R09			& $0.31\pm0.06$	& $1.97\pm0.02^{\dagger}$ \\
B13			& $0.32\pm0.05$	& $2.27\pm0.06$ \\
\hline
\end{tabular}
\tablecomments{The results of the fit of $\Gamma = m$log$_{10}$\lamedd$+c$ given different values of \kbol, where column (1) lists the \kbol\ used, column (2) lists the slope, $m$, and column (3) lists the offset, $c$, both with 1-$\sigma$ uncertainties. The last three lines give the results from samples of unobscured AGN for comparison. $^{\dagger}$R09 calculate their linear fit with log$_{10}$\lamedd$=-1$ as their reference point, rather than 0 as we have done here, which corresponds to $c=2.28$ with log$_{10}$\lamedd$=0$ as the reference point.}
\end{center}
\end{table}

Finally we investigate other sources of bolometric luminosity that are independent of the X-ray measurements. These usually come from fitting spectral energy distributions of the AGN from optical to mid-infrared wavelengths. Table \ref{tab_lbol} lists these, along with the corresponding \kbol\ for the given \lx\ of the AGN. No independent \lbol\ measurement could be found for NGC 4945 or IC 2560. Firstly this shows that \kbol\ for our sample shows a large spread of $\sim6-120$, with a median value of 33, an average of 45 and a standard deviation of 44. Although the sample is small, this appears systematically higher than the results from \cite{lusso12}. If indeed \kbol\ is dependent on \lamedd, a systematically higher \kbol\ for these CTAGN may imply a systematically higher \lamedd\ given the same \lx\ for unobscured AGN. Alternatively, the mid infrared from which the \lbol\ values have been estimated may include contributions from star-formation that have not been underestimated in the SED fitting. Indeed, NGC~3079, which stands out in out sample with \kbol=120, has a known nuclear starburst.
 
When using these values of \lbol\ instead of \lbol\ derived from \lx\ (and retaining the values from \lx\ for NGC 4945 and IC 2560), we obtain $\Gamma=( 0.28\pm 0.06)$log$_{10}$\lamedd$+( 2.18\pm 0.05)$ from \chisq\ minimization and $\Gamma=( 0.31\pm 0.20)$log$_{10}$\lamedd$+( 2.21\pm 0.19)$ (intrinsic scatter of 0.24$\pm$ 0.24) from the Bayesian method. Despite the large spread in \kbol\ and it being apparent systematically higher than the value of 10 that we use, the slope of the relationship is within 1-$\sigma$ of the \kbol=10 result, and the results from unobscured AGN. 

\begin{table}
\centering
\caption{Investigating bolometric luminosities from the literature.}
\label{tab_lbol}
\begin{center}
\begin{tabular}{l c c c c}
\hline
AGN Name & \lbol\ & \kbol\ & Method & Reference \\
(1) & (2) & (3) & (4) & (5)\\
\hline
NGC 1068& 45.0& 44&Flux integration&a\\
NGC 1194& 44.7& 91&MIR&b\\
NGC 2273& 44.0&  8.7&Flux integration&a\\
NGC 3079& 43.6&120&MIR&b\\
NGC 3393& 44.9& 33&[Ne V]&c\\
NGC 4388& 43.4&  5.9&MIR&d\\
Circinus& 43.6& 13&MIR&e\\

\hline
\end{tabular}
\tablecomments{Column (1) lists the AGN name for which a bolometric luminosity could be found in the literature, column (2) list the logarithm of \lbol\ in \ergs, column (3) lists the corresponding X-ray bolometric correction, \kbol, given this literature \lbol\ and \lx\ from Table \ref{tab_sample}, column (4) lists the method used for estimating \lbol\ and column (5) gives the reference, where a - \cite{woo02}, b - \cite{gruppioni16}, c - \cite{koss15}, d - \cite{ralmeida11}, e - \cite{moorwood96}}
\end{center}
\end{table}

\subsection{How does the X-ray spectral modeling affect our results?}

Our results on $\Gamma$ and \lamedd\ for the heavily obscured megamasers are also dependent on the X-ray torus model used to model the spectrum. As described above, for our analysis we have compiled results from both {\tt mytorus} and {\tt torus}. For most sources, the authors fitted both models and presented the best fitting case. We test to what extent this choice may have affected our results by compiling the results from {\tt mytorus} only, since this model was more commonly used. When using {\tt mytorus}, however, two sources produced ambiguous results. For NGC~2273 the model produced two degenerate results, one where $\Gamma>2.44$ and one where $\Gamma<1.4$ (A. Masini, private communication). Since there is ambiguity we do not include this source in our analysis with {\tt mytorus}. For IC~2560, {\tt mytorus} reaches the upper limit in both $\Gamma$ (2.5) and \nh\ ($10^{25}$ \cmsq). Since the {\tt torus} model indicates that \nh$>10^{26}$ \cmsq\ in this source, beyond the range of {\tt mytorus}, the result from {\tt mytorus} may not be reliable and thus we also do not include this source in our analysis with {\tt mytorus}. This is in agreement with \cite{balokovic14} where more detailed spectral modeling of this source is presented. Given then the 11 remaining data points, we carry out the same analysis as above, with \kbol=10 yielding $\Gamma=( 0.31\pm 0.07)$log$_{10}$\lamedd$+( 2.35\pm 0.06)$ from \chisq\ minimization and $\Gamma=( 0.36\pm 0.21)$log$_{10}$\lamedd$+( 2.31\pm 0.25)$ (intrinsic scatter of 0.27$\pm$ 0.26) from the Bayesian method. While the slope of this relationship is slightly steeper than for the mixed sample, it is statistically consistent within the uncertainties, as is the offset, indicating the the choice of torus model does not affect our result significantly.

Lastly, we discuss the two sources that we excluded from our analysis, NGC~1386 and NGC~2960. As with NGC~2273, fits with {\tt mytorus} to NGC~1386 yielded two degenerate solutions, one with a low $\Gamma$ and one with a high $\Gamma$ \citep{masini16}. With the {\tt torus} model, \cite{brightman15} found $\Gamma=2.9\pm0.4$, which is very high for any value of \lamedd. Similarly, the {\tt torus} model yields $\Gamma=2.6\pm0.4$ for NGC~2960 (\cite{masini16} fix $\Gamma$ in their fit with {\tt mytorus}). It is not clear whether these very high values of $\Gamma$ are related to the low-count nature of their spectra, or if they represent true outliers in the $\Gamma$-\lamedd\ relationship for the megamasers. Only longer exposures with \nustar\ will solve this question. We estimate that around $\sim2000$ counts at minimum in \nustar\ FPMA and FPMB are required for an accurate determination of $\Gamma$ in CTAGN, where NGC~1386 and NGC~2960 have less than 1000.

\section{Discussion and Conclusions}
\label{sec_disc}

From the above analysis, albeit with a small sample, we conclude that the low-mass, heavily obscured megamaser AGN are statistically consistent with the higher mass, unobscured AGN through the $\Gamma$-\lamedd\ relationship, where $\Gamma=( 0.41\pm 0.18)$log$_{10}$\lamedd$+( 2.38\pm 0.20)$. This result has the following implications. 

Firstly, the agreement indicates that the X-ray torus models effectively recover the intrinsic AGN parameters given sensitive broadband X-ray spectral data, despite the Compton-thick levels of absorption present in the megamaser systems. This has important value for results on heavily obscured AGN from \nustar\ and future X-ray missions with hard X-ray sensitivity, and is particularly timely due to several new compilations of X-ray torus models \citep[e.g.][]{liu14,furui16}.

Secondly, considering the low-\mbh\ nature of the megamasers (\mbh$\approx10^{6}-10^{7}$ \msol), our results imply that the connection between the accretion-disk emission, parameterized by \lamedd, and the physical state of the X-ray emitting corona, parameterized by $\Gamma$, is constant over $\sim3$ orders of magnitude in black hole mass, \mbh$\approx10^{6}-10^{9}$~\msol\ and $\sim2$ orders of magnitude in \lamedd\ ($\approx0.01-1$). A correlation between $\Gamma$ and \lamedd\ is also found in X-ray binaries, where \mbh$\approx10$ \msol, for the same \lamedd\ regime \citep[e.g.][]{yang15}, although the slope of the relationship, 0.58$\pm$0.01, is different to what we find for our AGN. Other results in the mass range we have investigated, selected on their small broadline widths \citep[i.e. narrow-line Seyfert 1s,][]{ai11} or their X-ray variability \citep{kamizasa12,ho16} do not find a significant correlation between $\Gamma$ and \lamedd. However these results cover a smaller range in \lamedd\ ($\sim1$ order of magnitude) and with larger uncertainties on their black hole mass estimates, which may be the reason they did not detect the correlation. 

Thirdly, since the megamaser AGN are edge-on systems, our results imply the lack of strong orientation effects when viewing the corona, which is assumed to be viewed more face-on in the unobscured AGN. This gives broad support to the AGN unification scheme \citep{antonucci93,urry95} and theoretical modeling of the AGN disk-corona system that predicts that the spectral shape in the X-ray band is insensitive to the viewing angle \citep{you12, xu15}. One potential caveat to this, however, is that there are known misalignments between the outer edge of the accretion disk/inner edge of the torus, where the masing occurs, and the alignment of the inner disk/corona in some objects. See \cite{lawrence10} for a discussion. Furthermore, orientation dependence of spectral properties in X-ray binaries have been reported by \cite{heil15}, where a difference of $\Delta\Gamma\approx0.17$ between low and high inclination systems is reported. Such a difference would manifest itself in our results in the offset of the $\Gamma$-\lamedd\ relationship. However, due to our small sample size, the statistical uncertainty in the offset is larger than 0.17, and furthermore we have shown there are systematic uncertainties in the offset due to estimation of \lbol. Therefore differences at this level are currently undetectable.

Finally, although there is significant scatter in the relationship between \lamedd\ and $\Gamma$ for the heavily obscured AGN, as there is for unobscured AGN, there is potential for $\Gamma$ to be used to give an indication of \lamedd\ in heavily obscured AGN where none would otherwise exist. For example, from the low $\Gamma$ of 1.75 measured by \cite{puccetti16} in the \nustar\ spectrum of the highly absorbed system NGC~6240 supported the low accretion rate inferred in the source. However, this method should be restricted to large samples in order to reduce the effect of the intrinsic scatter. In the future, {\it eROSITA} \citep{merloni12} will measure the $0.5-10$ keV spectra of millions of AGN over the whole sky and {\it ATHENA} \citep{nandra13} will have the ability to carry out the measurements required for AGN at high redshift in deep surveys, inferring \lamedd\ distributions from large samples of AGN over a wide redshift range.

\acknowledgments

This work was supported under NASA Contract No. NNG08FD60C, and made use of data from the {\it NuSTAR} mission, a project led by the California Institute of Technology, managed by the Jet Propulsion Laboratory, and funded by the National Aeronautics and Space Administration. We thank the {\it NuSTAR} Operations, Software and Calibration teams for support with the execution and analysis of these observations. Furthermore, this research has made use of the NASA/IPAC Extragalactic Database (NED) which is operated by the Jet Propulsion Laboratory, California Institute of Technology, under contract with the National Aeronautics and Space Administration. A.M. and A.C. acknowledge support from the ASI/INAF grant I/037/12/0-011/13. P.G. acknowledges funding from STFC (ST/J003697/2). M.\,Balokovi\'{c}. acknowledges support from NASA Headquarters under the NASA Earth and Space Science Fellowship Program, grant NNX14AQ07H.

{\it Facilities:} \facility{\nustar}

\bibliography{bibdesk.bib}

\end{document}